\documentclass[prl,twocolumn,showpacs,amsmath,amssymb,floatfix]{revtex4}
\usepackage{mathpazo,times,graphicx}
\usepackage[colorlinks=true,citecolor=blue]{hyperref}
\usepackage{cleveref}
\allowdisplaybreaks
\def\reftitle#1,{\relax}
\def\etalii#1,{\textit{et al.},}

\def\be{\begin{equation}}
\def\ee{\end{equation}}
\def\bse{\begin{subequations}}
\def\ese{\end{subequations}}

\makeatletter
\def\overl@ss#1#2{\vcenter{\offinterlineskip
        \ialign{$\m@th#1\hfil##\hfil$\crcr#2\crcr<\crcr } }}
\def\gl{\mathrel{\mathpalette\overl@ss>}}
\def\e{\mathop{\rm e}\nolimits}

\def\diag{\mathop{\rm diag}\nolimits}

\def\Real{\mathbb{R}}
\def\Complex{\mathbb{C}}

\def\Im{\mathop{\rm Im}\nolimits}
\def\@#1{{\mathbf{#1}}}
\def\_#1{{\mathsf{#1}}}
\let\~=\tilde
\let\^=\hat
\let\<=\langle
\let\>=\rangle

\let\eref=\eqref
\makeatother

\def\re{\mathrm{re}}
\def\im{\mathrm{im}}
\def\today{September 9, 2015}

\begin{document}
\title{Universal nature of the nonlinear stage of modulational instability}
\author{Gino Biondini$^{1,2,*}$ and Dionyssios Mantzavinos$^2$}
\affiliation{$^1$ State University of New York at Buffalo, Department of Physics, Buffalo, NY 14260, USA
\\
$^2$ State University of New York at Buffalo, Department of Mathematics, Buffalo, NY 14260, USA}
\date{\small\today}
\begin{abstract}
We characterize the nonlinear stage of modulational instability (MI) by studying the long-time asymptotics of 
focusing nonlinear Schr\"odinger (NLS) equation on the infinite line 
with initial conditions that tend to constant values at infinity.
Asymptotically in time, the spatial domain divides into three regions:
a far left field and a far right field, in which the solution is approximately equal to its initial value,
and a central region in which the solution has oscillatory behavior
and is described by slow modulations of the periodic traveling wave 
solutions of the focusing NLS equation.
These results demonstrate that the asymptotic stage of MI is universal, since 
the long-time behavior of a large class of perturbations is described by the same asymptotic state.
\end{abstract}
\maketitle

\paragraph{Introduction.}
Modulational instability (MI) 
--- i.e., the instability of a constant background to long wavelength perturbations --- 
is one of the most ubiquitous phenomena in nonlinear science (e.g., see \cite{zakharovostrovsky} and references therein). 
The effect, 
which is known as Benjamin-Feir instability in the context of deep water waves \cite{benjaminfeir},
has been known since the 1960's, 
but has received renewed attention in recent years, 
and was also linked to the formation of rogue waves in optical media \cite{naturephys,solli} and in the open sea \cite{onoratoosborne}. 

In many cases, the dynamics of systems affected by MI is 
governed by the one-dimensional focusing nonlinear Schr\"odinger (NLS) equation, which models the evolution of weakly nonlinear dispersive wave packets 
in such diverse fields as water waves, plasmas, optics
and Bose-Einstein condensates.
One can therefore study the initial (i.e., linear) stage of MI by linearizing the NLS equation around the constant background. 
One easily sees that all Fourier modes below a certain threshold are unstable, and the corresponding perturbations grow exponentially.
However, the linearization ceases to be valid as soon as perturbations have become comparable with the background.
A natural question is then what happens at this point, which is referred to as \textit{the nonlinear stage of MI}.
Surprisingly, 
a precise characterization of the nonlinear stage of MI for generic, finite-energy perturbations 
has remained by and large an open problem for the last fifty years.

The NLS equation is a completely integrable system \cite{zakharovshabat1972}, 
and admits an infinite number of conservation laws and exact $N$-soliton solutions for arbitrary $N$, 
describing the elastic interaction of solitons \cite{zakharovshabat1972}.
By analogy with the case of localized initial conditions, 
a natural conjecture was that MI is therefore mediated by solitons \cite{gelash,zakharovgelash}.
%
The initial value problem (IVP) for the NLS equation can be solved via the inverse scattering transform (IST).
In particular, the IST for the focusing NLS equation with zero boundary conditions (ZBC) at infinity (i.e., localized disturbances)
was done in \cite{zakharovshabat1972}, 
and the IST for the defocusing NLS equation with nonzero boundary conditions (NZBC, i.e., solutions that tend to finite value at infinity) 
was done in \cite{zakharovshabat1973}.
But only partial results \cite{kuznetsov,ma,garnier}
were available for the focusing NLS equation with NZBC until recently, 
in \cite{biondinikovacic}, we developed a complete IST for this case.
(Recall that the IST for systems with NZBC is notoriously more challenging, 
and the IVP for the vector NLS with NZBC was also only solved recently \cite{prinari,kbk2015}.)
In \cite{biondinifagerstrom} we then used the IST to study MI
by computing the spectrum of the scattering problem for simple classes of perturbations of a constant background.
In particular, we showed that \textit{there are classes of perturbations for which no solitons are present.
Thus, since all generic perturbations of the constant background are linearly unstable, 
solitons cannot be the mechanism that mediates the MI},
which contradicts a recent conjecture \cite{zakharovgelash}.
Instead, in \cite{biondinifagerstrom} we identified the instability mechanism within the context of the IST, 
by showing that the instability comes from the continuous spectrum of the scattering problem associated with the NLS equation
(see below for further details).

In this Letter we use the framework developed in \cite{biondinikovacic} to characterize the nonlinear stage of MI.
We do so by studying the long-time asymptotic behavior of localized perturbation of the constant background. 
We show that, generically, the long-time asymptotics of modulationally unstable fields on the whole line 
displays universal behavior, and decomposes the $xt$-plane into two plane wave regions
--- in each of which the solution is approximately equal to the background up to a phase ---
separated by a central region in which the leading order behavior is described by a slowly modulated traveling wave solution.

\medskip
\paragraph{The NLS equation and MI.}

We write the focusing NLS equation as
\vspace*{-1ex}
\be
iq_t + q_{xx} + 2(|q|^2 - q_o^2) q = 0\,,
\label{e:NLS}
\ee
where 
$q(x,t)$ 
represents the complex envelope of a quasi-monochromatic, weakly nonlinear dispersive wave packet,
and the physical meaning of the variables $x$ and $t$ depends on the physical context. 
(E.g., in optics, $t$ represents propagation distance while $x$ is a retarded time.) 
Here $q_o = |q_\pm|>0$ is the background amplitude, and
the NZBC satisfied by the field are
\vspace*{-1.4ex}
\be
q_\pm = \lim_{x\to\pm\infty}q(x,t)
\label{e:NZBC}
\ee
The term $-2q_o^2q$ has been added to Eq.~\eref{e:NLS} so that $q_\pm$ are independent of time,
and can be removed by a trivial gauge transformation.

The constant background solution is simply $q_s(x,t) = q_o$.  
Linearizing Eq.~\eref{e:NLS} around this solution,
one finds that all Fourier modes 
with $|\zeta|<2q_o$ (where $\zeta$ is the Fourier variable) are unstable,
and that the growth rate is $\gamma(\zeta) = |\zeta|\sqrt{4q_o^2-\zeta^2}$.
Below we will use the IST for Eq.~\eref{e:NLS} with the NZBC~\eref{e:NZBC},
that was developed in~\cite{biondinikovacic},
slightly reformulated in a way that is more convenient for the present purposes.

Recall that the NLS Eq.~\eref{e:NLS} is the zero-curvature condition $X_t - T_x + [X,T] = 0$ of the matrix 
Lax pair $\phi_x = X\phi$ and $\phi_t = T\phi$, with 
$X = ik\sigma_3 + Q$ and $T = -i(2k^2+q_o^2-|q|^2-Q_x)\,\sigma_3 -2kQ$, where 
$\sigma_3 = \diag(1,-1)$ is the third Pauli matrix,  and
\vspace*{-1ex}
\be
Q(x,t) = \begin{pmatrix} 0 &q \\ - q^* & 0 \end{pmatrix}.
\ee
As usual, the first half of the Lax pair is referred to as the scattering problem and $q(x,t)$ as the potential,
and the direct problem in the IST consists in determining the scattering data (i.e., reflection coefficient, discrete eigenvalues and norming constants)
from the IC.
This is done through the Jost eigenfunctions $\phi_\pm(x,t,k)$, 
which are the simultaneous matrix solutions of both parts of the Lax pair which reduce to plane waves, namely,
$\phi_\pm(x,t,k) = E_\pm(k)\e^{i\theta(x,t,k)\sigma_3} + o(1)$ as $x\to\pm\infty$,
where $\pm i\lambda$ and $E_\pm(k) = I + i/(k+\lambda)\,\sigma_3 Q_\pm$ 
are respectively the eigenvalues and corresponding eigenvector matrices of $X_\pm = \lim_{x\to\pm\infty}X$, 
with 
$\lambda(k) = (k^2+q_o)^{1/2}$ 
and 
$\theta(x,t,k) = \lambda x-\omega t$, 
and where
$\omega(k) = 2k\lambda$.
These Jost eigenfunctions, which are the nonlinearization of the Fourier modes, 
are defined for $k\in\Complex$ such that $\lambda(k)\in\Real$, which defines the continuous spectrum
$\Sigma = \Real\cup i[-q_o,q_o]$, see Fig.~1(left).
The scattering relation $\phi_-(x,t,k) = \phi_+(x,t,k)A(k)$ defines the scattering matrix $A(k)$ for $k\in\Sigma$,
and the corresponding reflection coefficient is $r(k) = -a_{21}/a_{22}$.
The zeros of $a_{11}(k)$ and $a_{22}(k)$ define the discrete spectrum of the problem, which leads to solitons.
As usual, time evolution within IST is trivial.  In particular, 
with the above normalization of the Jost eigenfunctions, all the scattering data are independent of time.

The focusing NLS Eq.~\eref{e:NLS} with the NZBC~\eref{e:NZBC} possesses a rich family of soliton solutions 
\cite{kuznetsov,peregrine,akhmediev,tajiriwatanabe}, 
classified according to the possible placements of the discrete eigenvalue \cite{biondinikovacic}.
In particular, the so-called Akhmediev breathers provide a good representation for the growth of seeded perturbations \cite{ablowitzherbst1990,henderson1999}.
Importantly, however, Akhmediev breathers are periodic in space, and therefore possess infinite energy.  
Hence they cannot describe the asymptotic state of localized (i.e., finite-energy) perturbations of the constant background.
Moreover, as mentioned earlier, 
generic perturbations of the constant background exist for which no discrete spectrum (and thus no solitons) is present.
Instead, 
\textit{the key to describe the asymptotic stage of MI lies in the continuous spectrum}.
Indeed, as we showed in \cite{biondinifagerstrom},
$\omega(k)$ is purely imaginary for $k\in i[-q_o,q_o]$,
and \textit{the Jost solutions for $k\in i[-q_o,q_o]$ are precisely the nonlinearization of the unstable Fourier modes}.
In fact, even their growth rate is the same, modulo the usual rescaling.

The inverse problem in the IST consists in reconstructing the solution $q(x,t)$ of the NLS equation from the scattering data, 
and is formulated in terms of a Riemann-Hilbert problem, 
namely the problem of reconstructing the meromorphic matrix  
$M(x,t,k)$ defined as 
$M(x,t,k) = (\phi_{+,1}/a_{22},\phi_{-,2})\,\e^{-i\theta\sigma_3}$ for $k\in\Complex^+\setminus i[0,q_o]$ and
$M(x,t,k) = (\phi_{-,1},\phi_{+,2}/a_{11})\,\e^{-i\theta\sigma_3}$ for $k\in\Complex^-\setminus i[-q_o,0]$,
where $\Complex^\pm = \{k\in\Complex:\Im k\gl0\}$
and $\phi_{\pm,j}$ for $j=1,2$ denote the columns of $\phi_\pm$.
This is done by using the scattering relation and symmetries to obtain a jump condition
$M^+(x,t,k) = M^-(x,t,k)V(x,t,k)$ for $k\in\Sigma$, where 
superscripts $\pm$ denote projection from the left/right of the contour $\Sigma$
(oriented rightward along the real $k$-axis and upward along the segment $i[-q_o,q_o]$).
Explicitly, 
\begin{gather*}
V(x,t,k) = 
\begin{cases}
\begin{pmatrix} 1+|r|^2 & r^* \e^{2i\theta} \\ r \e^{-2i\theta} & 1 \end{pmatrix},\quad k\in\Real\,,\\
\displaystyle \frac{iq_o}{k-\lambda}\begin{pmatrix} -r^*\e^{2i\theta} & 1 \\ 1+|r|^2 & -  r \e^{-2i\theta} \end{pmatrix},\quad k\in i[0,q_o]\,,
\end{cases}
\end{gather*}
plus a symmetric expression for $k\in i[-q_o,0]$.
Note that $\det M(x,t,k) = 1$ for $k\in\Complex\setminus\Sigma$ and $M(x,t,k) \to I$ as $k\to\infty$.
The solution of the NLS is recovered via the usual reconstruction formula 
$q(x,t) = -2i\lim_{k\to\infty}k M_{12}$.
The signature of MI in the inverse problem is the exponentially growing entries of $V(x,t,k)$ 
for $k\in i[-q_o,q_o]$
through the time dependence of $\theta(x,t,k)$.

\begin{figure}[b!]
\includegraphics[width=0.225\textwidth]{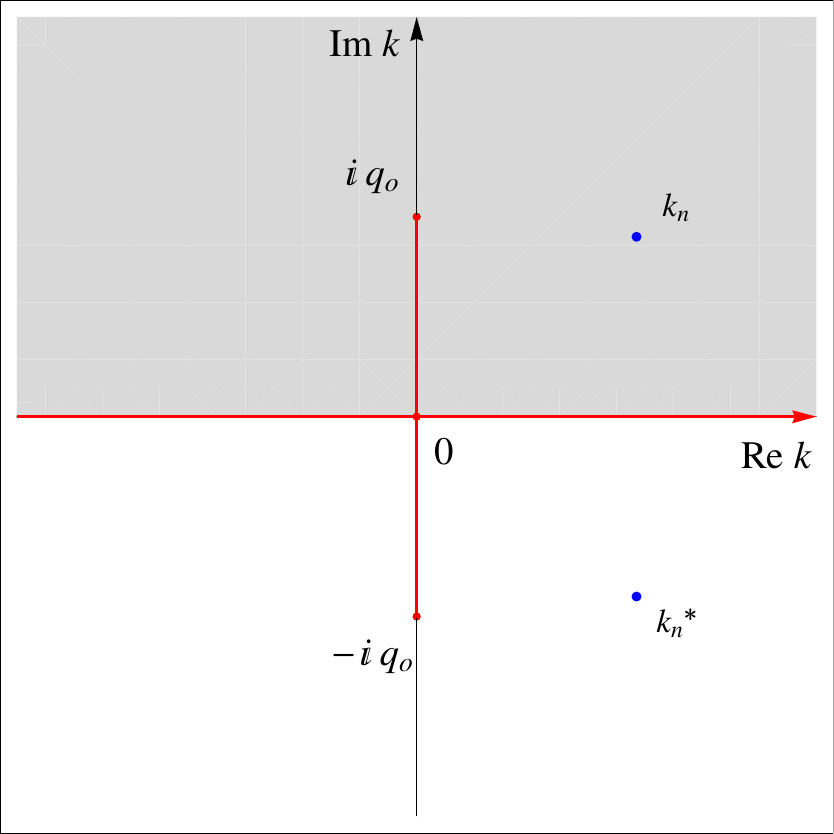}~~~
\includegraphics[width=0.225\textwidth]{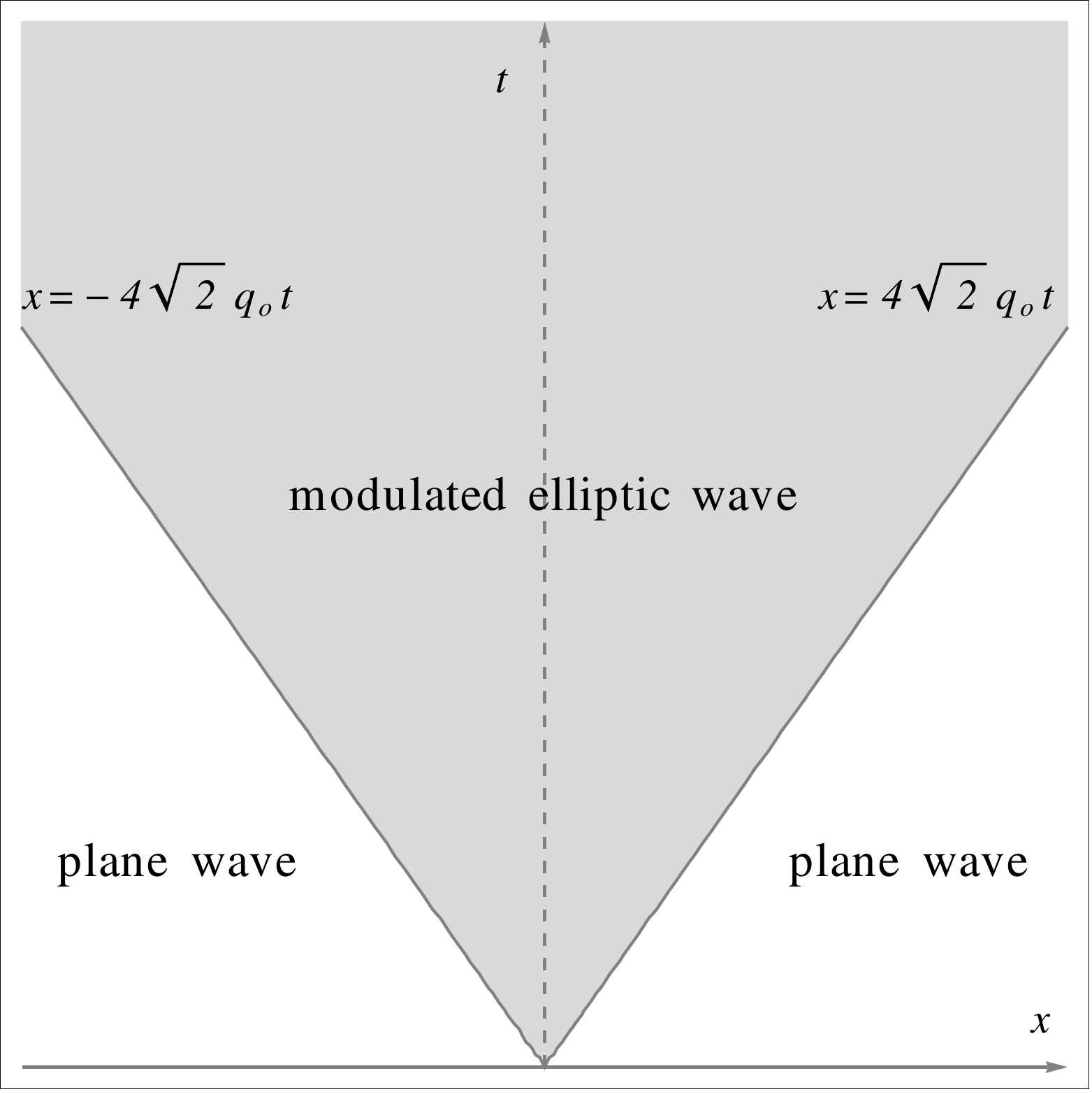}
\begin{caption}
{Left: the spectral $k$-plane, showing the continuous spectrum $\Sigma$ (red lines), 
the region where $\Im\lambda>0$ (gray)
and a discrete eigenvalue together with its symmetric counterpart (blue and brown, respectively).
Right: the asymptotic regime for the $xt$-plane, showing the decomposition into two plane wave regions (white)
and the modulated elliptic wave (gray).}
\end{caption}
\label{f1}
\end{figure}

\medskip
\paragraph{Long-time asymptotics of finite-energy perturbations.}

We now study the asymptotic state of MI for generic, finite-energy perturbations of a constant
background. As mentioned earlier, we do so by computing the long-time asymptotics of the solutions of 
the focusing NLS equation with NZBC.
As a concrete example we consider box-like perturbations with $q(x,0) = q_o$ for $|x|>L$ and
$q(x,0) = b\e^{i\beta}$ for $|x|<L$, in which case 
$r(k) = \e^{2i\lambda L}[(b\cos\beta-q_o)k-ib\lambda\sin\beta]/[\lambda\mu\cot(2L\mu)-i(k^2+q_ob\cos\beta)]$,
with $\mu = \sqrt{k^2+b^2}$.
The calculations, however, apply to all localized perturbations
such that the corresponding reflection coefficient
has a small region of analyticity around the continuous spectrum.

Recall that for linear evolution equations one computes the asymptotics of the solution as $t\to\infty$
via stationary phase or steepest descent
by looking along lines $x=\xi t$ with $\xi$ fixed
\footnote{The point $x=0$ appears to be special because, 
in the far field approximation of the dynamics arising from localized perturbations, everything seems to arise from the origin,
just like in the far-field asymptotics for linear problems \cite{whitham}.}.
In this far field approximation, the solution essentially becomes the Fourier transform of the initial condition,
modulated by the similarity variable $\xi$ and evaluated at the critical points of the problem \cite{whitham}.
In the nonlinear case, instead, one must use the IST. 

The long-time asymptotics of solutions of the NLS equation with ZBC was computed through various approaches in~\cite{ablowitzsegur1976,zakharovmanakov}.
Those results, however, do not apply in our case.
Here we used the more general nonlinearization of the steepest descent method, 
namely the Deift-Zhou (DZ) method for oscillatory RHP
\cite{deiftzhou}.
The essence of the DZ method is to perform appropriate deformations of the 
matrices and contours of the RHP to ``peel away'' the complicated time dependence 
and reduce oneself to an asymptotic, model problem that can be solved exactly.
Much like for its linear counterpart (i.e., the steepest descent method), however, 
the specific application of the DZ method is highly problem-dependent.
In our case, the DZ method allows us to modify the RHP so that growing jumps are replaced by 
bounded ones 
This result is not a fluke, but is a typical feature of the DZ method \cite{buckingham,boutet}.

\begin{figure}[t!]
\vglue1ex
\includegraphics[width=0.12\textwidth]{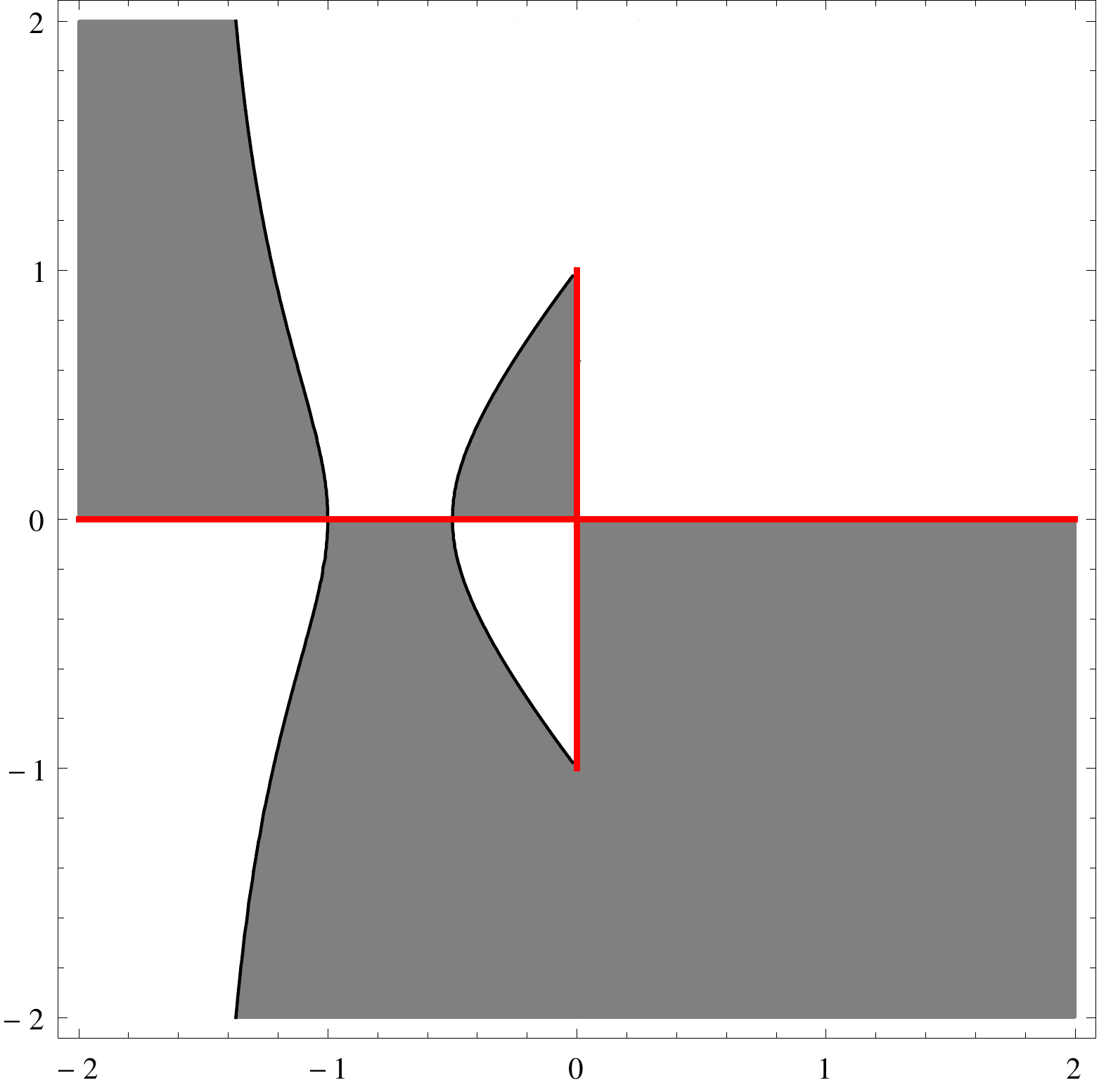}~%
\includegraphics[width=0.12\textwidth]{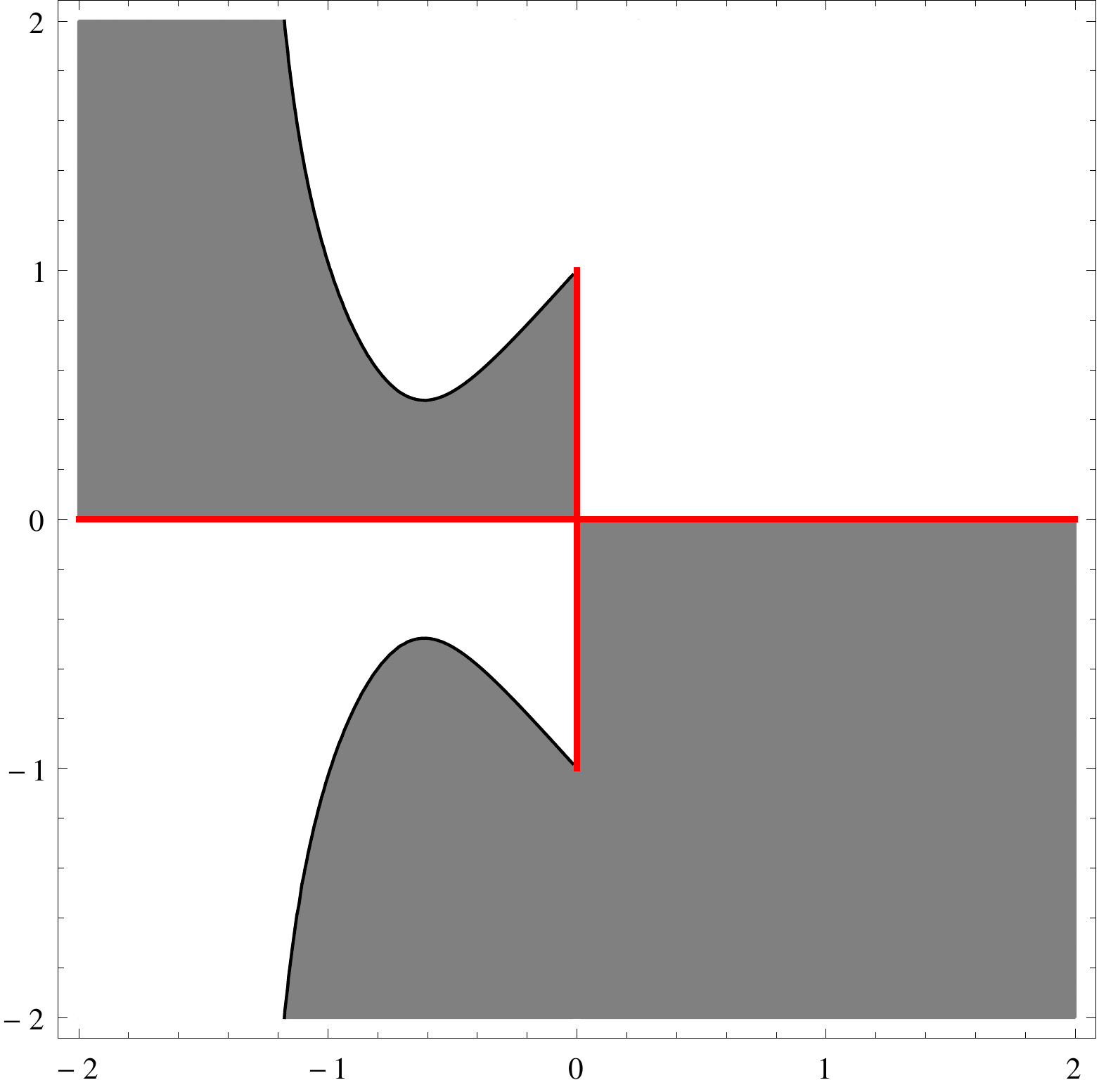}~%
\includegraphics[width=0.12\textwidth]{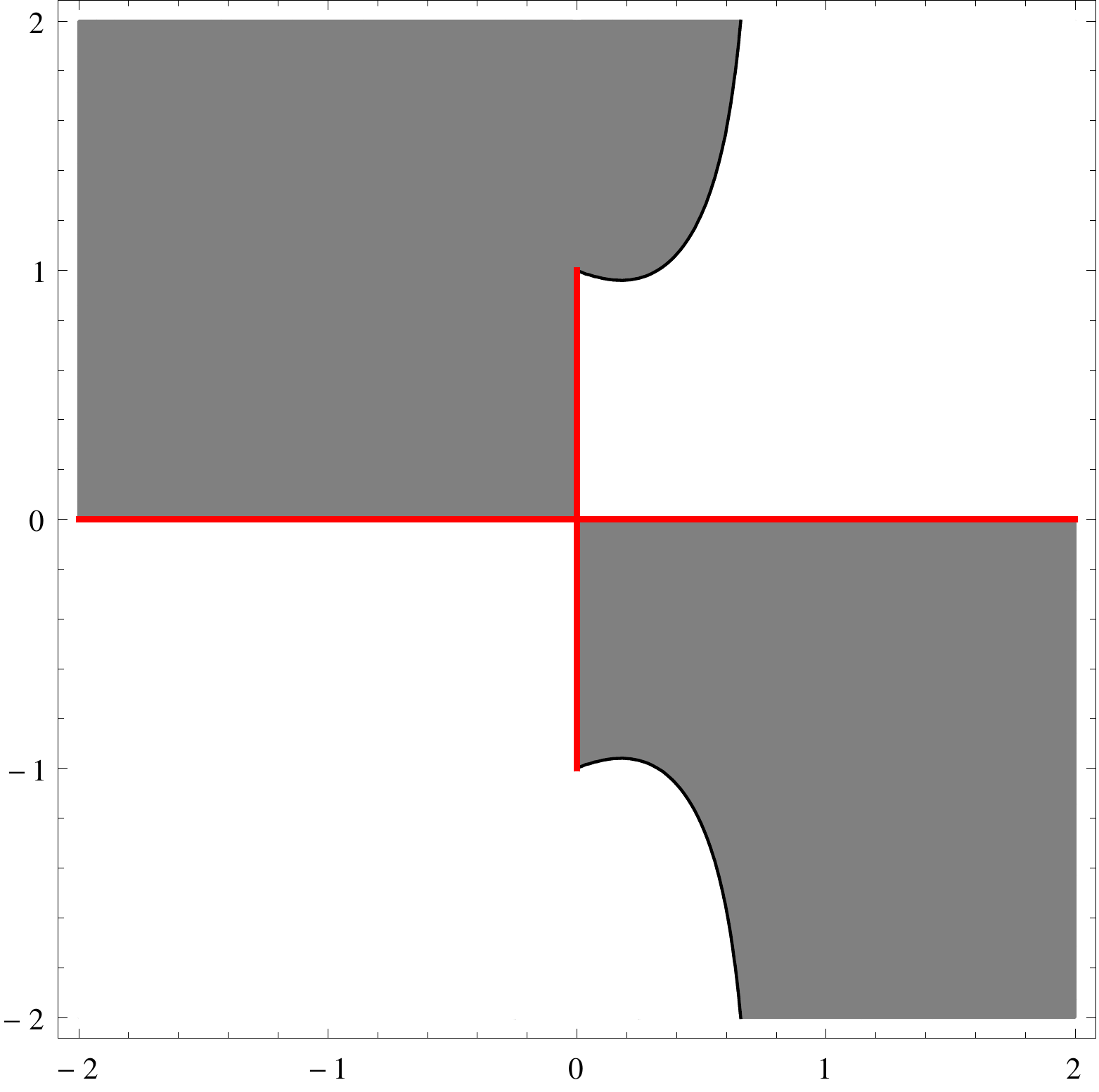}~%
\includegraphics[width=0.12\textwidth]{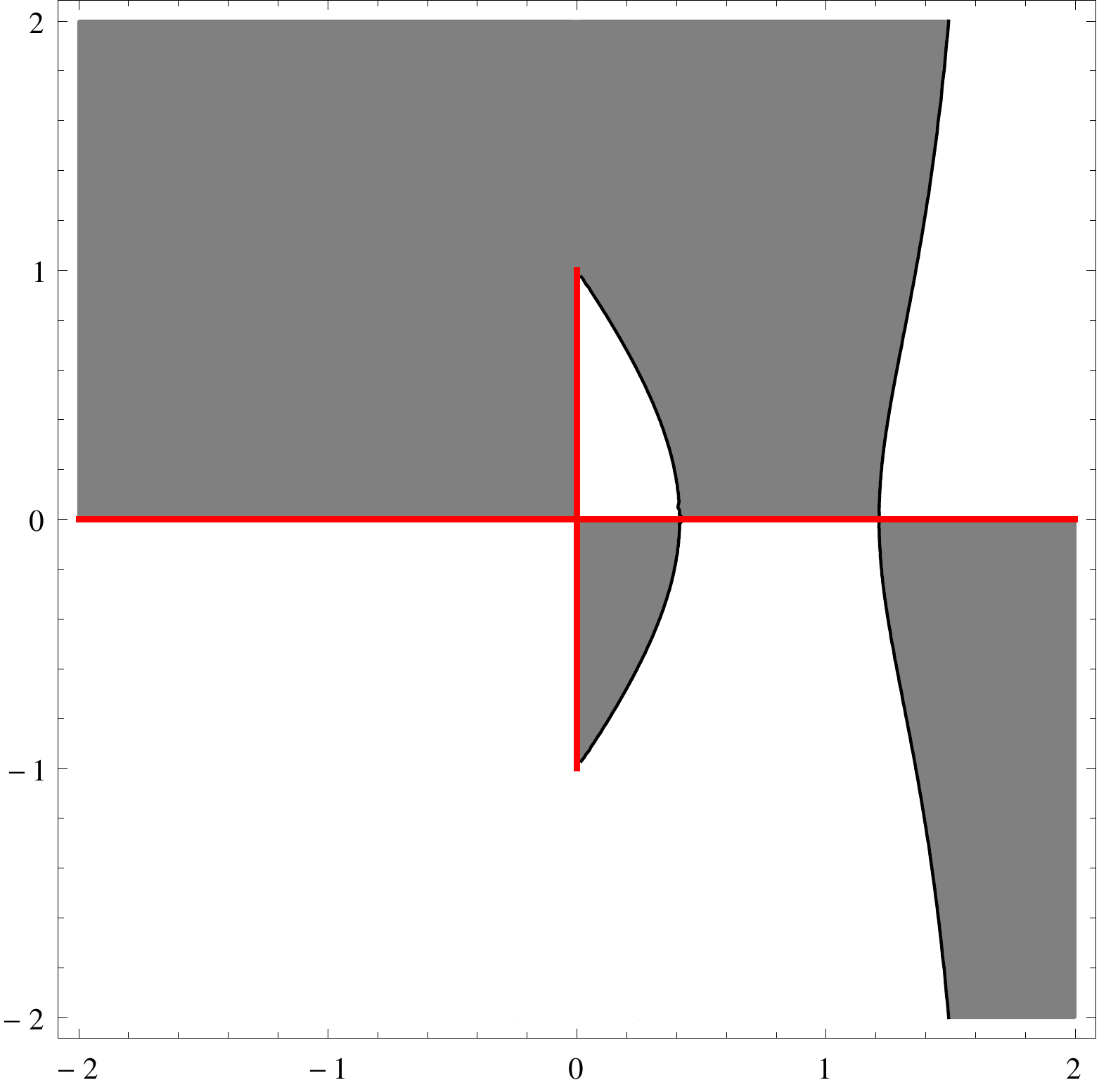}%
\begin{caption}
{The sign structure of $\Im\theta$ in the complex $k$-plane for various values of $\xi$ for $q_o=1$:
(i)~$\xi = -6$, corresponding to $x< - \xi_*t$;
(ii)~$\xi = -5.2$, corresponding to $\xi_*t<x<0$;
(iii)~$\xi = 3$, corresponding to $0<x< \xi_*t$;
(iv)~$\xi = 6.5$, corresponding to $x > \xi_*t$.
Gray: $\Im\theta>0$. White: $\Im\theta<0$.}
\end{caption}
\vglue-2.5ex
\end{figure}

\medskip
\paragraph{Asymptotic stage of MI.}

As the implementation of the DZ method is complicated, the details will be reported elsewhere.
On the other hand, the main results can summarized in a straightforward way.
The key piece of information in the deformation of the RHP is the sign structure
of $\Im\theta = \Im[\lambda(\xi-2k)]t$ as a function of $k$ for $\xi$ fixed.
Let $\xi_* = 4\sqrt2 q_o$.
For $|x|>\xi_*t$, there are two real stationary points in the complex $k$-plane.
This situation correspond to the first and fourth plot of Fig.~2.
For $|x|<\xi_*t$, there are two complex conjugate stationary points in the complex $k$-plane. 
This situation correspond to the second and third plot of Fig.~2.

Each of the four cases in Fig.~2 requires a different deformation in the DZ method.
Correspondingly, the $xt$-plane is divided into three regions, 
as illustrated in the bifurcation diagram in Fig.~1(right).
Specifically:
(i)~The range $x < - \xi_* t < 0$ is the left far field, plane wave region.
Here $|q(x,t)| = q_o + O(1/t^{1/2})$ as $t\to\infty$.
Apart from a nonlinear contribution to the phase, the behavior is similar to the linear case.
(ii)~The range $- \xi_* t < x < \xi_* t$ is an oscillation region.
Here $q(x,t) = q_\mathrm{asymp}(x,t) + O(1/t^{1/2})$, 
the asymptotic solution being a modulated traveling wave (elliptic) solution. 
This is the most interesting region, and is
described in some detail below.
(iii)~The range $x > \xi_* t > 0$ is a right far field, plane wave region.
Here, $|q(x,t)| = q_o + O(1/t^{1/2})$, as $t\to\infty$, similarly to region~(i).

The detailed expression of the solution in each region will be reported in a separate publication.
The kind of results described above are not unprecedented, however.  
Indeed, bifurcation diagrams dividing the long-time asymptotic behavior of solutions of the focusing NLS equation
into regions of various genus were obtained in different contexts in \cite{boutet,buckingham}. 
What is different here, however, is the physical setting,
the specific results and their physical interpretation.

\medskip
\paragraph{The modulated traveling wave region.}

We focus on the range $0<x<\xi_*t$. (The expression for the solution in the range $-\xi_*t<x<0$ is similar.)
The leading-order solution in this region is expressed in terms of elliptic functions,
and represents a slow modulation of the traveling wave (periodic) solutions
of the focusing NLS equation
\footnote{This solution was first studied in \cite{el,kamchatnov} in the context of Whitham theory,
but neither work studied the evolution of generic initial conditions.}. 
In particular,
\vspace*{-1ex}
\begin{multline}
|q_{\mathrm{asymp}}(x,t)|^2 = (q_o + \alpha_\im)^2
\\
- 4q_o\alpha_\im\,\mathop{\rm sn}\nolimits^2[C(x-2\alpha_\re t-X);m],
\end{multline}
where $m = 4q_o\alpha_\im/C^2$ is the elliptic parameter,
$C=\sqrt{\alpha_\re^2 + (q_o + \alpha_\im)^2}$,
and the slowly varying offset $X$ is explicitly determined by the reflection coefficient. 
The four points $\pm iq_o$ and $\alpha_\pm = \alpha_\re \pm i\alpha_\im$ are the branch points 
associated with the elliptic solutions of the focusing NLS equation \cite{el,kamchatnov};
$\alpha_\pm$ are slowly varying functions of~$\xi$, 
determined via a single, implicit equation that can be easily solved numerically.
The slowly varying wavenumber, velocity and period are respectively $\alpha_\re$, $2\alpha_\re$ and $2K(m)/C$ 
\footnote{Here $K(m)$ is the complete elliptic integral of the first kind. 
Since the wave is nonlinear, the wavenumber and period are not related by a simple proportionality
relation as for harmonic waves.}.
In particular, $\alpha \to 1/\sqrt2$ as $x\to\xi_* t$ and $\alpha\to iq_o$ as $x\to0$.
The first limit corresponds to the boundary between the genus-1 region and the plane wave region,
in which case $m\to0$ and the solution reduces to a constant.
In the second limit, $m\to1$, corresponding to the solitonic limit of the elliptic solution.

The universal profile of the solution amplitude in the oscillation region
(neglecting for simplicity the $\xi$-dependent effect of the reflection coefficient)
is shown in Fig.~3 at two different values of time.
The envelope of the solution (dashed lines), given by $q_o\pm\alpha_\im$, is time-independent,
and depends only on~$\xi$. 
Conversely, the oscillating structure is slowly varying in the\break $xt$-frame.
The boundary between the oscillation region and the plane wave regions 
can be understood within the context of Whitham modulation theory \cite{el,kamchatnov}.

\begin{figure}[t!]
\centerline{\begin{minipage}{0.24\textwidth}%
\includegraphics[height=0.4\textwidth,width=0.85\textwidth]{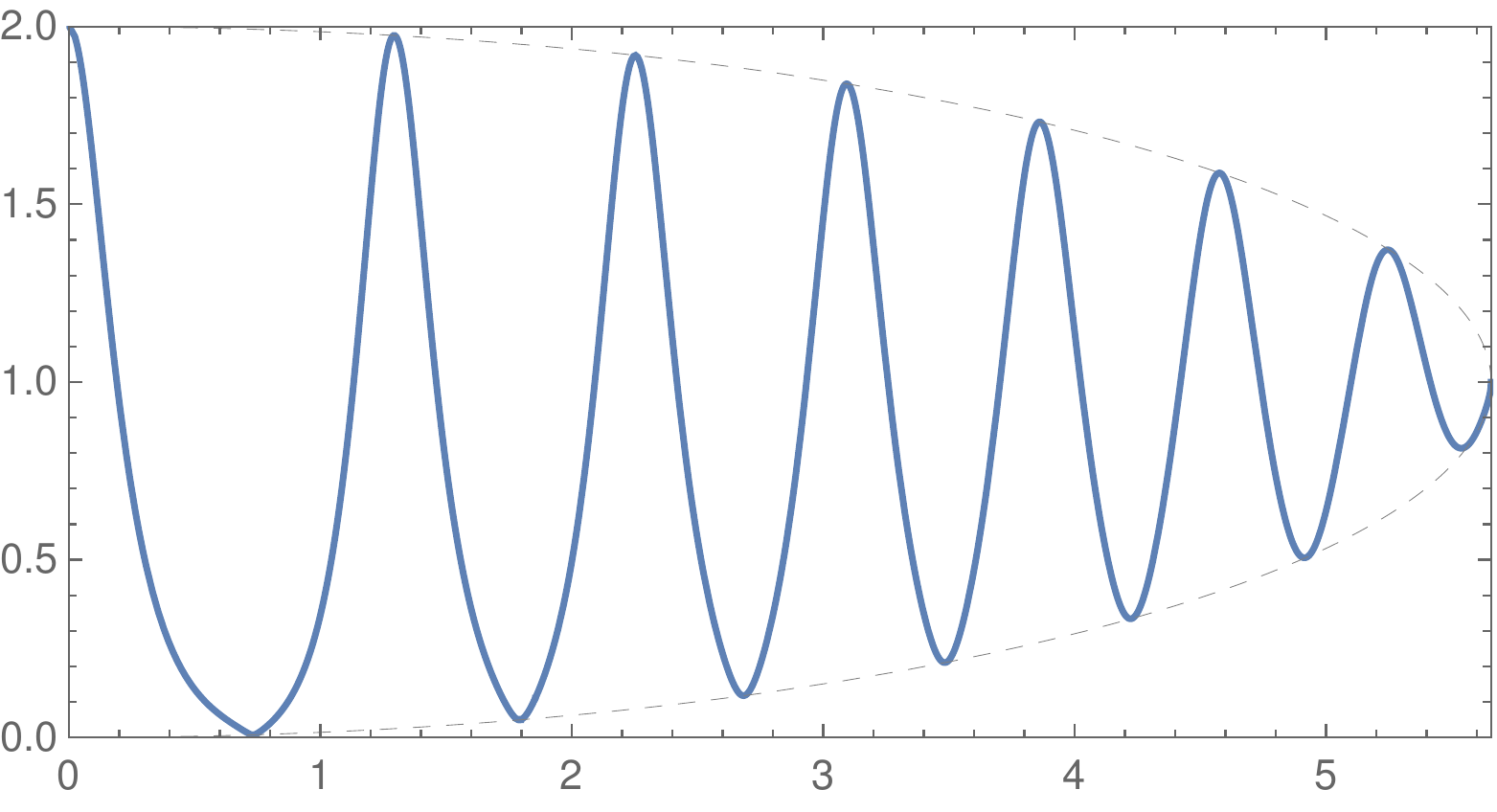}\\
\includegraphics[height=0.4\textwidth,width=0.85\textwidth]{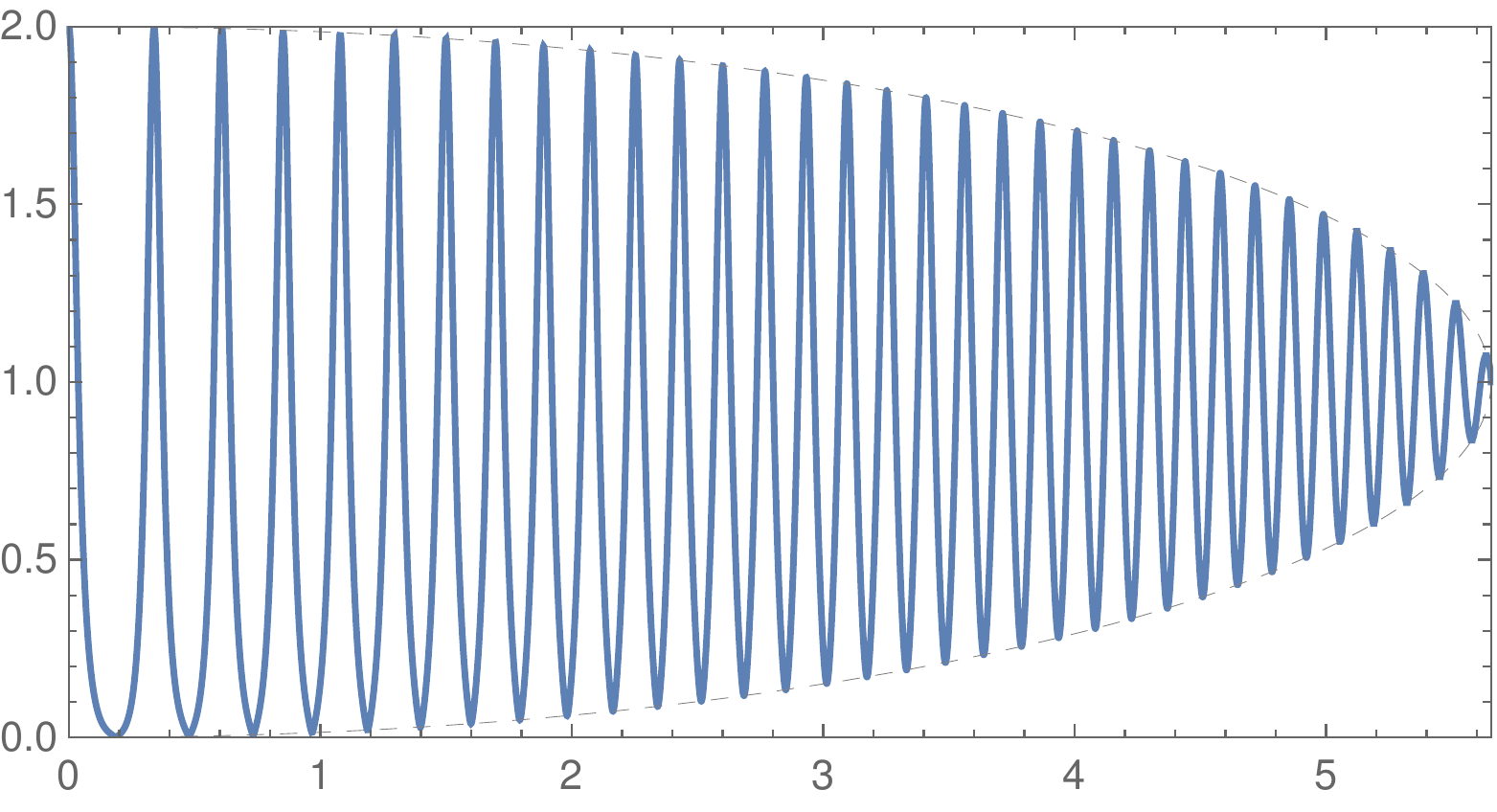}%
\end{minipage}\kern-1em
\begin{minipage}{0.295\textwidth}%
\includegraphics[width=\textwidth]{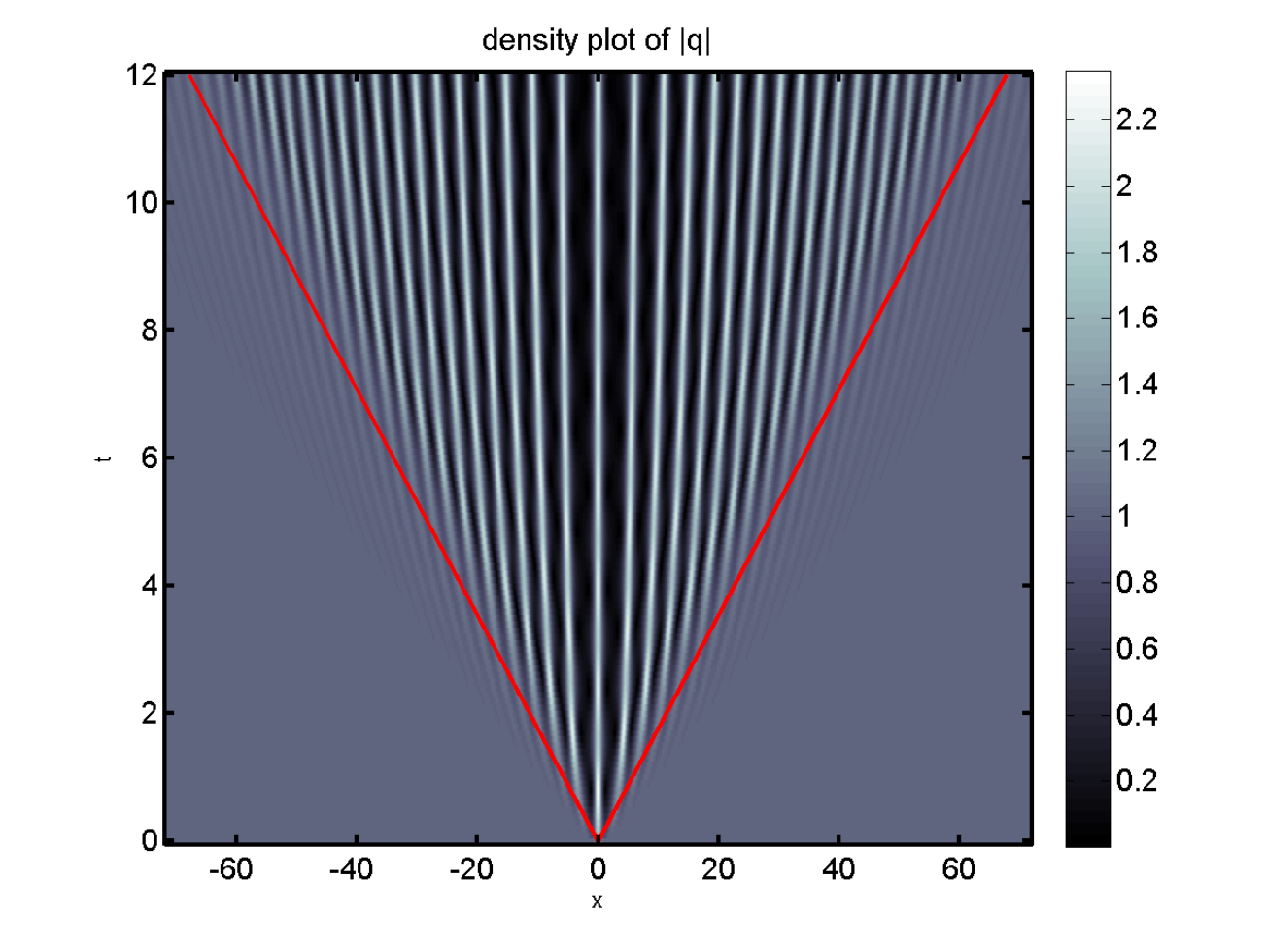}%
\end{minipage}}
\kern-\medskipamount
\begin{caption}
{Left: The asymptotic solution $|q(x,t)|$ (vertical axis) in the oscillation region as a function
of $\xi=x/t$ (horizontal axis) at different values of time for $q_o=1$. 
Top:~$t=4$.
Bottom:~$t=20$.
Also shown (dashed lines) is the time-independent envelope of the solution.
Right: 
Density plot from numerical simulations of Eq.~\eref{e:NLS} with a small Gaussian perturbation of the constant background.
The red lines show the analytically predicted boundaries $x= \pm4\sqrt2q_ot$.
}
\end{caption}
\vglue-1ex
\end{figure}

\medskip
\paragraph{Discussion.}
We have computed the long-time asymptotic behavior of a large class of perturbations to a constant background 
in a modulationally unstable medium for which no discrete spectrum is present.
We emphasize the broad nature of our results. 
The initial conditions of the problem only determine a slowly varying offset
for the elliptic solution via the reflection coefficient,
whereas the structure of the solution as a modulated elliptic wave is independent of it.
In this sense, \textit{the asymptotic stage of MI is universal}.

Since the NLS equation has a wide range of applicability, 
from nonlinear optics to deep water waves, acoustics, plasmas and Bose-Einstein condensates,
we expect that the results of this work will apply to all of the above physical contexts.
The results also have potential 
connections to the phenomena of rogue waves \cite{naturephys,solli} 
and integrable turbulence \cite{zakharov2009}.

The results of this work should be compared to those in the case of periodic boundary conditions (BC).
There, the instability is ascribed to the presence of homoclinic solutions \cite{prl71p2683}.
The initial stage of MI was studied in \cite{Trillo1} with a 3-mode model.
But the machinery used to solve the IVP in the periodic case 
(namely, the theory of finite-genus solutions \cite{itskotlyarov,maablowitz})
is very different from that used here.
The limiting process from the periodic case to the infinite line is highly nontrivial, 
and has not been properly understood yet. 
Most importantly, the physics in the two cases is different.  
For example:
(i) In the periodic case there is an amplitude threshold below which no instability occurs,
whereas no such threshold exists on the infinite line. 
(ii) In the periodic case, radiation cannot escape to infinity, 
and therefore it is doubtful that a long-time asymptotic state even exists.

The above results can also be compared to the semiclassical limit 
of the focusing NLS equation with ZBC \cite{KMM}.  
The study of that scenario requires more sophisticated analysis, and the results are also more complicated
\footnote{Although there is a similar bifurcation from plane waves to modulated genus-2 oscillations in the case of ZBC, 
the genus-2 solution also breaks along a certain caustic curve in the $xt$-plane, 
with numerical evidence indicating the presence of regions 
of higher and higher genus in the limit $\hbar\to0$.}.  
Moreover, numerical simulations of the semiclassical case become more and more sensitive to round-off error as $\hbar\to0$ \cite{prl71p2683}.  
(Essentially, the IVP becomes ill-posed in that limit.)
In contrast, the present case does not appear to be as sensitive.  
As a result, there are no fundamental obstacles to the possibility that the behavior described in this work could observed experimentally.
The robustness of the analytical predictions is confirmed in Fig.~3, which shows a numerical simulations of Eq.~\eref{e:NLS}
with a small Gaussian perturbation of the constant background.

Semiclassical limits and long-time asymptotics problems 
are often studied using Whitham theory \cite{whitham}.
But the Whitham equations for the focusing NLS equation are elliptic, and therefore the corresponding IVP is ill-posed. 
This is well known in the case of ZBC (e.g., see \cite{KMM}), 
and it remains true in the case of NZBC. 
While special solutions to the Whitham equations also exist in the focusing case 
\cite{el,kamchatnov},
it should be clear that the IST-related methods used here are the only way to study the nonlinear stage of MI
for generic perturbations of the constant background.
Indeed, we see no obstacles to generalizing the present calculations to include the presence of discrete eigenvalues,
which will allow for the first time a study of the interactions between solitons and radiation in 
modulationally unstable media.

\input references

\end{document}